# Phonons and Anomalous Thermal Expansion Behaviour of H$_2$O and D$_2$O ice I$h$

M. K. Gupta[1], R. Mittal[1,2], Baltej Singh[1,2], S. K. Mishra[1], D. T. Adroja[3,4], A. D. Fortes[3] and S. L. Chaplot[1,2]

[1]Solid State Physics Division, Bhabha Atomic Research Centre, Mumbai, 400085, India
[2]Homi Bhabha National Institute, Anushaktinagar, Mumbai 400094, India
[3]ISIS Facility, Rutherford Appleton Laboratory, Chilton, Didcot, Oxon OX11 0QX, UK
[4]Highly Correlated Matter Research Group, Department of Physics, University of Johannesburg, Auckland Park 2006, South Africa

In order to identify and quantitatively analyze the anharmonicity of phonons relevant to the anomalous thermal expansion in the I$h$ phase of ice, we performed neutron inelastic scattering measurements of the phonon spectrum as a function of pressure up to 1 kbar at 225 K in deuterated ice (D$_2$O), and as a function of temperature over 10-225 K at ambient pressure in both H$_2$O and D$_2$O ice. We also performed density functional theory calculations of the lattice dynamics. The anomalous expansion is quantitatively reproduced from the analysis of the neutron data as well as from the ab-initio calculations. Further, the ab-initio calculations are used to visualize the nature of anharmonic phonons across a large part of the Brillouin zone. We find that the negative thermal expansion below 60 K in the hexagonal plane is due to anharmonic librational motion of the hexagonal rings of the ice molecules, and that along the hexagonal axis originates from the transverse vibrations of the hexagonal layers.

Understanding the behaviour of ice over a broad range of thermodynamic conditions is of great importance in the field of Earth and planetary sciences as well as in fundamental physics and chemistry [1-21]. Ice is known to exhibit polymorphic structures, depending on the environmental conditions[1]. It exists even in outer space, in comet clouds and interstellar grains[7-9]. The known phase diagram of ice is extremely rich [1]. The hydrogen bonds in ice are weak enough to be bent, stretched, or shortened substantially under pressure, which explains various crystalline structures that have been identified so far.

Ice has many anomalous properties and perhaps a most complex phase diagram. Ordinary ice (I$h$) crystallizes in an open structure which provides ample possibilities for rearrangements with pressure and temperature. The region of stability of ice I$h$ has been extensively studied[1, 12-14] both by experiments as well as theory. Among all the known crystalline ice phases, ice (I$h$ phase) exhibits anomalous thermal expansion [12, 13] behaviour. It exhibits negative thermal expansion (NTE) at temperatures below ~ 60 K. As temperature increases above 60 K, it shows a large positive thermal expansion coefficient. The understanding of the mechanism of the anomalous thermal expansion involves an understanding of the underlying dynamics of hydrogen-bonded water molecules in ice and their behaviour on compression. Here we focus on one of the peculiarities, namely, the anomalous expansion behaviour with temperature in ice I$h$. The thermodynamic stability of various phases of ice including the role of vibrational anharmonicity[2, 14, 19, 20, 22-33] have been studied by various groups. *Koza et al* [15-17] reported extensive neutron and x-ray scattering experiments dedicated to the study of the dynamics of some high-density crystalline and amorphous ice structures. Spectroscopic studies have been reported using dielectric, infrared (IR), Raman and neutron measurements[21, 34-37]. Strassle, *et al*[14] have measured the pressure dependence of low energy phonon dispersion in deuterated ice (D$_2$O) to understand the pressure induced amorphization and NTE behaviour. The measurements along the high symmetry directions in the Brillouin zone show softening of low energy modes of about 5-7 meV. The phonon density of states of H$_2$O and D$_2$O ice in I$h$ phase [38] has been measured at 15 K. Edgar et al [20] theoretically studied the significance of phonon anharmonicity on the stability of hexagonal ice over the stability of cubic phase. The isotope effect of H and O on ice volumes has been studied [4]. Various theoretical methods[4] have been used to calculate the thermal expansion in ice. Others studies on ice addressed the proton hopping/disorder and diffusion in ice using dielectric and quasielastic neutron scattering method [21, 39].

Since the mechanism of the anomalous thermal expansion involves anharmonic phonons over the entire Brillouin zone, it is necessary to study the complete phonon spectrum. Similarly, there is a need to investigate the anharmonicity and phonon eigenvectors across the entire Brillouin zone by ab-initio density functional theory. We have characterised the pressure and temperature dependence of the phonon spectrum using the inelastic neutron scattering technique and performed extensive ab-initio lattice dynamics calculations of H$_2$O and D$_2$O ice in the I$h$ phase. We are able to derive the thermal expansion at low temperature from the experimental phonon spectrum. The results bring out the nature of specific anharmonic soft phonon modes that show anomalous behaviour as a function of pressure and temperature, which leads to the anomalous thermal expansion behaviour of ice. Details of the experiments and calculations are given in supplementary information[40].

We have measured (Fig. 1) the phonon spectra in I$h$ phase of D$_2$O ice at 0 kbar, 0.3 kbar and 1.0 kbar (at 225 K) well below the phase transition pressure of about 2 kbar at 225 K where ice I$h$ transforms to ice II. The low incident neutron energy of 15 meV ensured the good resolution for the low-energy phonon required to obtain the Grüneisen parameters of phonon modes. The spectrum up to 4 meV clearly shows the Debye like behaviour and a peak at ~7 meV.

The basic chracteristics of the low energy peak is in good agreement with previous measurements [38]. On increase of pressure, the low energy phonon modes soften and shift to lower energies. This gives rise to negative Grüneisen parameters for these modes resulting in negative volume thermal expansion at low temperature. The present results on the pressure dependence of the phonon density of states along selected directions in the Brillouin zone are consistent with an earlier report of phonon dispersion relation on ice I$h$ performed by Strassle et al [14], which shows the softening of low energy phonon modes with pressure. Using standard methods [41, 42] described in the supplementary material[40], the pressure dependence of phonon spectrum is used to obtain the Grüneisen parameter, which is further used to estimate (inset in Fig. 2) the contribution of phonons of energy E to the volume thermal expansion coefficient at 50 K. It can be seen that phonons of energy around 6 meV contribute the most to the observed NTE behaviour. As shown below, this observation is in excellent agreement with our ab-initio calculations.

We have measured (Fig 3) the inelastic neutron spectraat temperatures ranging from 10 K to 225 K for $H_2O$ and $D_2O$ in the I$h$ phase. The measurements are performed with two incident neutron energies of 15 meV and 55 meV. The 15 meV incident neutrons give inelastic scattering data upto about 12 meV with very good resolution ($\Delta E \sim 0.6$ meV at elastic line); with 55 meV incident neutrons we can obtain inelastic scattering data up to 45 meV of moderate resolution ($\Delta E \sim 2.45$ meV). The spectra show sharp peaks at about 7, 20, 28 and 38 meV in both the compounds. The low energy modes around 6 meV show softening with increase in temperature which can be clearly seen in both the compounds. This is characteristic of the presence of anhamonic phonon modes. The measured inelastic spectrum shows broadening of various peaks at higher temperature, whicharises due to the anharmonic nature of the phonons as well as an increase in H/D disorder with temperaure. At high temperature (~225K) the signature of H/D disorder dominates and the sharp peak-like features of the spectrum disappear. This disorder behaviour of H/D plays an important role in the thermal expansion behaviour at higher temperaure.

The calculated neutron-weighted phonon density of states compared with the measurements is shown in Fig 3. The calculation agrees fairly well with the measurements at low temperature at 10 K. However, the peaks at 28 and 38 meV in the experimental data are slightly overestimated and underestimated respectively in the calculated spectra. The negative expansion behaviour occurs below 60 K and the contributions of high energy modes is not very significant at such low temperatures, which gives us confindence to understand the anamolous behaviour of $H_2O$ and $D_2O$ in the I$h$ phase with the current computational formalism. An examination of the calculated phonon eigenvectors shows that the modes below 40 meV are essentially the translational modes of $H_2O$/$D_2O$ molecules. The calculated neutron cross-section weighted partial phonon density of states shows (Fig. 3) that at low energies, the experimental neutron spectra in $H_2O$ ice is dominated by H atoms due to the large incoherent cross section of H. However in the experimental neutron spectra of $D_2O$ ice, both the D and O atoms have comparable contributions.

We obtained the softening of low energy modes below 12 meV on compression from the calculations [40]. Table I shows the calculated Grüneisen parameters of a few selected phonon modes of about 6 meV. Since lowering of the phonon frequency would enhance the entropy, these phonon modes(<12 meV) act to reduce the volume of the system. However, the energy of other higher-energy modes increases on compression which would decrease the entropy. These modes will not favor volume contraction and instead work to expand the lattice. The competition between these two regimes of phonon modes and their population govern the overall thermal expansion behaviour. At low temperature (<60 K) only the low energy modes are populated significantly. The other high-energy modes are not sufficiently populated to compete with these low energy modes. Hence $D_2O$/$H_2O$ ice I$h$ shows negative thermal expansion behaviour at low temperature(Fig. 4). Further, at high temperature above 60 K the high energy modes above 12 meV start populating significantly and dominate, which leads to positive expansion behaviour.

Since the hexagonal structure of the I$h$ phase is anisotropic, we have calculated the anisotropic thermal expansion using the anisotropic Grüneisen parameters of phonons and the elastic compliance matrix as derived from the ab-initio DFT [40]. The experimental measurements[12, 13] of thermal expansion behaviour show that magnitude of NTE below 40 K in $D_2O$ ice is slightly more in comparison to $H_2O$ ice, while at high temperature the magnitude of positive thermal expansion behaviour in $D_2O$ ice is larger than that in $H_2O$ ice (Fig. 4). This can be understood from the calculated partial and total phonon spectrum. The negative thermal expansion is contributed by low energy phonon modes below 12 meV. The partial and total phonon spectrum looks very similar in both the cases. However due to mass effect the lowest peak at about 7 meV in density of states in $H_2O$ is shifted to slightly lower energies at about 6.5 meV. The nature of Grüneisen parameters show a negative sign below 12 meV and afterward it is positive. Hence the expansion behaviour along both the 'a-' and 'c-' axis shows negative expansion behaviour at low temperature (<60K) and after that it shows positive expansion behaviour. The calculated fractional change in the lattice parameters with temperature shows excellent agreement with the available measurements[12] at temperature below 100 K (Fig 4). At temperature around 200 K, the calculated thermal expansion behaviour is underestimated in comparison to the experimental data. There could be two factor which may be responsible for this underestimation of thermal expansion coefficient at high temperature. First, the elastic compressibility (inverse of bulk modulus) of ice is known to exhibit large temperature dependence and is found to show an increases of about 30% on increase of temperature from 50 K to 250 K [43]. Which means that some of the elastic compliance component would increase singnificantly with temperature and would be responsible for increase in positive expansion behaviour at high temperature. Another reason may be due to the fact that entropy contribution in I$h$ phase of ice $H_2O$ and $D_2O$ ice would have contribution from phonon vibrations as well as from H/D disorder. The H/D positional disorder would increase with lengthening of the H···O bond ie. with increase in volume. The shorter covalent O-H bond is much stronger and does not change much with temperature. Hence in order to increase the entropy, the disorder in the the H/D position will increase and result in increase of unit cell volume significantly with increase of temperature. The present formalism of calculation of thermal expansion behaviour does not account for both these effects, hence may lead to underestimation of calculated volume thermal expansion at higher temperature. The magnitude of positive expansion in $D_2O$ (Fig 4) is larger than that in $H_2O$, which is due to the fact that high energy phonon modes which contribute to positive expansion are shifted towards lower energy side, hence they are excited at lower temperature and add to the extra postive expansion in comparison to $H_2O$.

In order to estimate the phonons which contribute mainly to positive and negative expansion behaviour,the calculated pressure dependence of phonon energies are used to estimate the thermal expansion coefficient as a function of phonon energy (Fig. 2 and Fig. S2[40]) averaged over the entire Brillouin zone at 50 K. Interestingly we find that phonon modes below 12 meV give rise to negative linear expansion coefficient. The high-energy phonons(in the range from 12-40 meV) contribute to very small positive contribution resulting in overall NTE behaviour. However with increase of temperature to 100 K, the negative contribution by phonon of energy below 12 meV does not change significanly. The additional large positive contribution from high energy modes is enhanced singnificantly and also contributes to the total thermal expansion resulting in overall large positive expansion.The ab-initio calculations and observations from high pressure experiments (inset in Fig. 2) at 50 K are found to be in excellent agreement.

As discussed above low energy phonon modes below 12 meV contribute to negative expansion in both the I$h$ phase of ice $D_2O$ and $H_2O$. In Table I, we have given the Grüneisen parameters of a few selected phonon modes relevent to NTE behaviour in the compound. However the nature of the dynamics .i.e. the atomic motion pattern for such modes have not been known. In order to visualize these dynamics we have drawn the atomic displacement pattern of one representative phonon mode which contributesto NTE (Fig 5).

The displacement pattern of the K-point phonon mode of 4.7 meV energy is shown in Fig 5; this manifests itself as a rotationalal motion of the hexagonal rings and a distortion of the ring structure, which acts to reduce the area of the rings and thereby give rise to NTE in the a-b plane. The hexagonal rings, arranged into sheets, are stacked along the c- direction. These cooperative rotation of the hexagonal rings in one plane influences, due to polyhedral connectivity of the H bonds, the interplanar separation. Another mode (Fig 5) of similar energy ~4.5 meV at the A-point involves transverse vibrations of the hexagonal layers, which further contributes to NTE along the c-axis. These calculations show that it is the rotational dynamics of the hexagonal rings that leads to NTE in these ice. Animations of a few selected modes are given in supplementary material[40].

In this paper, we have quantitatively derived the anisotropic and anomalous thermal expansion in $H_2O$ and $D_2O$ ice in the I$h$ phase. This has also enabled to understand the quantitative difference in thermal expansion behaviour of the $H_2O$ and $D_2O$ ice. The inelastic neutron scattering measurements of the pressure dependence of phonon density of states of deuterated ice ($D_2O$) reveal a pronounced softening of low-energy anharmonic modes. The increase in broadening of the peaks in the temperature dependence of phonon spectra of both the protonated ($H_2O$) and deuterated ice ($D_2O$) indicates increase in disorder of H/D atoms with increasing temperature. Using the ab-initio density functional theory we have identified the nature of specific anharmonic phonons, namely, the librational motion of the hexagonal rings of the ice molecules and the transverse vibrations of the hexagonal layers that lead to the observed behaviour in ice.

**Acknowledgements**
S. L. Chaplot would like to thank the Department of Atomic Energy, India for the award of Raja Ramanna Fellowship. The use of ANUPAM super-computing facility at BARC is acknowledged. The authors also thank STFC, UK for the beam-time at ISIS and thank the pressure and furnace section technicians for their essential assistance in setting up the equipment.

TABLE I The calculated anisotropic Grüneisen parameter of a few phonons having large Grüneisen parameters in the Brillouin zone of an ordered analogue (space group P6$_3$cm) of ice I$h$. E and Γ are the energy and Grüneisen parameter respectively. The wave-vectors are Γ(0 0 0), A(0 0 0.5), K(.33 0.33 0), H(.33 0.33 0.5), M(0.5 0.5 0), and L (0.5 0.5 0.5)

| Wave vector | E(meV) | $\Gamma_a$ | $\Gamma_c$ | $\Gamma_v$ |
|---|---|---|---|---|
| | | $H_2O/D_2O$ | | |
| Γ | 5.6/5.3 | -2.0/-2.0 | -5.3/-5.2 | -3.1/-3.0 |
| A | 4.5/4.2 | -0.9/-0.9 | -3.7/-3.6 | -1.9/-1.8 |
| A | 6.8/6.5 | -3.0/-3.1 | -2.9/-2.8 | -3.0/-3.0 |
| K | 4.7/4.5 | -4.0/-3.9 | -3.0/-2.7 | -3.7/-3.5 |
| K | 5.2/4.9 | -2.9/-2.9 | -1.6/-1.4 | -2.5/-2.4 |
| K | 7.1/6.8 | -0.5/-0.5 | -0.7/0.7 | -0.6/0.6 |
| H | 7.3/6.9 | -1.6/-1.5 | -1.5/-1.5 | -1.6/-1.5 |
| H | 7.7/7.4 | -1.6/-1.7 | -1.6/-1.7 | -1.6/-1.7 |
| M | 5.3/5.0 | -3.9/-4.0 | -2.1/-2.2 | -3.4/-3.4 |
| M | 6.1/5.8 | -0.5/-0.5 | -0.2/-0.2 | -0.4/-0.4 |
| L | 7.5/7.2 | -1.6/-1.5 | -1.6/-1.7 | -1.6/-1.6 |
| L | 7.8/7.5 | -1.6/-1.6 | -1.4/-1.4 | -1.5/-1.5 |

FIG 1. (Color Online) The low energy part of experimental phonon spectra, in the I*h* phase of $D_2O$ ice, as a function of pressure. Additional data at 0.3 kbar is given in **Fig. S1[40].**

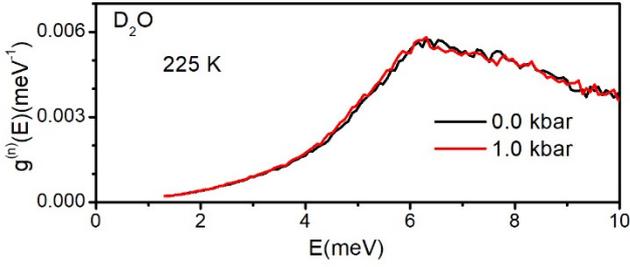

FIG 2. (Color Online) The ab-initio calculated contribution to the volume thermal expansion coefficient from phonons of energy E averaged over entire Brillouin zone at T=50K and 100 K. Inset shows the thermal expansion coefficient as a function of phonon energy in $D_2O$ ice as obtained from pressure dependence of neutron inelastic measurements and comparison with the calculation.

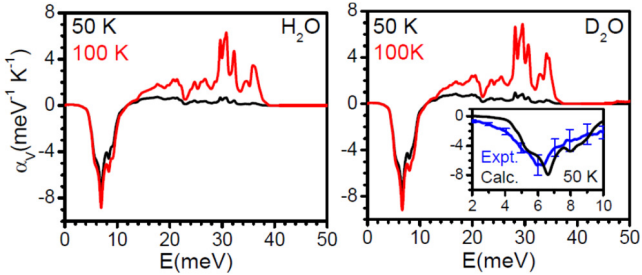

FIG 3. (Color Online) The low energy part of experimental phonon spectra, in the I*h* phase of $H_2O$ and $D_2O$ ice, as a function of temperature. The data are collected with two incident neutron energies of 15 meV and 55 meV. The calculated neutron weighted phonon spectra as well as partial phonon contributions of H/D and O atoms are also shown. For clarity, the experimental data is vertically shifted.

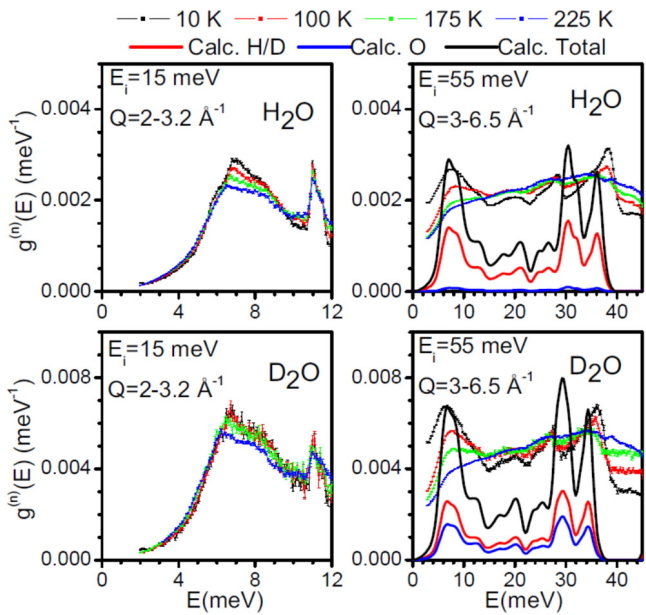

FIG 4. (Color Online) The calculated and measured[13] lattice parameter change ($l/l_{10}$; $l$= a,c) of ice with temperature. Inset in the left panel shows the calculated volume thermal expansion coefficient ($\alpha_V$) of $H_2O$ and $D_2O$ ice up to 60 K.

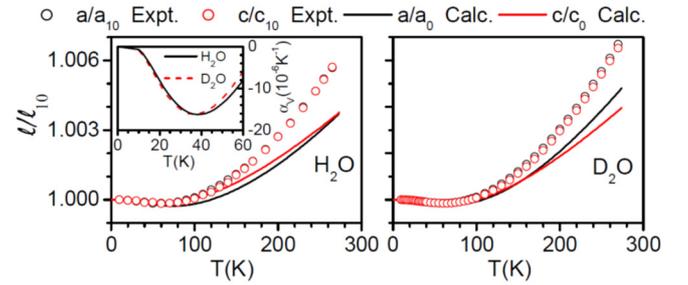

FIG 5. (Color Online) The atomic displacement pattern of selected phonon modes of $H_2O/D_2O$ in I*h* phase of ice. The number below the figure gives the phonon energies, $\Gamma_a$, $\Gamma_c$ and $\Gamma_V$ respectively. Key- H/D: green and O: red.

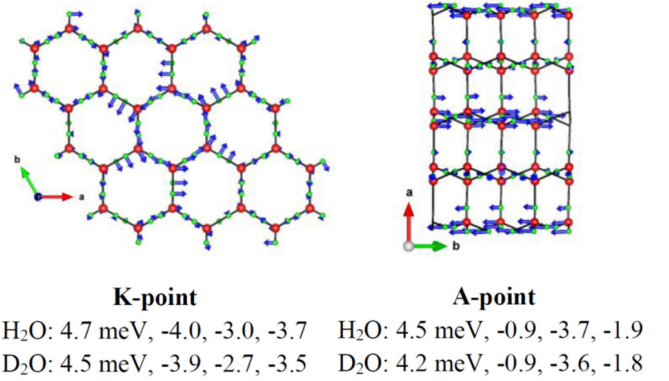

| K-point | A-point |
|---|---|
| $H_2O$: 4.7 meV, -4.0, -3.0, -3.7 | $H_2O$: 4.5 meV, -0.9, -3.7, -1.9 |
| $D_2O$: 4.5 meV, -3.9, -2.7, -3.5 | $D_2O$: 4.2 meV, -0.9, -3.6, -1.8 |

# Supplementary Materials

**Phonons and Anomalous Thermal Expansion Behaviour of $H_2O$ and $D_2O$ ice in I$h$ phase**

M. K. Gupta[1], R. Mittal[1,2], Baltej Singh[1,2], S. K. Mishra[1], D. T. Adroja[3,4], A. D. Fortes[3] and S. L. Chaplot[1,2]

[1]Solid State Physics Division, Bhabha Atomic Research Centre, Mumbai, 400085, India
[2]Homi Bhabha National Institute, Anushaktinagar, Mumbai 400094, India
[3]ISIS Facility, Rutherford Appleton Laboratory, Chilton, Didcot, Oxon OX11 0QX, UK
[4]Highly Correlated Matter Research Group, Department of Physics, University of Johannesburg, Auckland Park 2006, South Africa

**Experimental:**

The inelastic spectrum of $H_2O$ and $D_2O$ ice in the I$h$ phase were measured on the polycrystalline samples. The measurements were performed at MERLIN time-of-flight spectrometer at ISIS facility. The spectrometer is equipped with a large detector bank covering a wide range of about 20º to 120º of scattering angle. High pressure inelastic neutron scattering experiments have been carried out on polycrystalline sample of I$h$ phase of $D_2O$. The polycrystalline sample was compressed using argon gas in a gas pressure cell at different pressures of 0, 0.3 and 1 kbar at 225 K. An incident neutron energy of 15 meV is chosen, and the measurements are performed in the energy loss mode. The inelastic neutron scattering signal is corrected for the contributions from argon at the respective pressures, absorption from the sample, and for the empty cell.

For temperature dependence, we have used about 1 mm and 2 mm thick polycrystalline samples for measurement of the spectra from $H_2O$ and $D_2O$ ice respectively. We have performed measurements at 10, 100, 175 and 225 K using two different incident neutron energies of 15 meV and 55 meV respectively. In the incoherent one-phonon approximation[1, 2], the measured scattering function $S(Q,E)$, as observed in the neutron experiments, is related to the phonon density of states $g^{(n)}(E)$ as follows:

$$g^{(n)}(E) = A \left\langle \frac{e^{2W(Q)}}{Q^2} \frac{E}{n(E,T) + \frac{1}{2} \pm \frac{1}{2}} S(Q,E) \right\rangle \quad (1)$$

$$g^n(E) = B \sum_k \left\{ \frac{4\pi b_k^2}{m_k} \right\} g_k(E) \quad (2)$$

where the + or – signs correspond to energy loss or gain of the neutrons respectively and $n(E,T) = \left[\exp(E/k_B T) - 1\right]^{-1}$. $A$ and $B$ are normalization constants. $b_k$, $m_k$, and $g_k(E)$ are, respectively, the neutron scattering length, mass, and partial density of states of the $k^{th}$ atom in the unit cell. The quantity between <> represents suitable average over all $Q$ values at a given energy. $2W(Q)$ is the Debye-Waller factor averaged over all the atoms. The weighting factors $\frac{4\pi b_k^2}{m_k}$ in the units of barns/amu for H, D, and O are: 82.02, 3.82 and 0.2645 respectively. The values of neutron scattering lengths for various atoms can be found from Ref.[3].

**Computational Details**

The phonon calculations of $H_2O$ and $D_2O$ are performed in I$h$ phase using the ab-initio density functional theory as implemented in the VASP software [4, 5]. The compounds are known to exhibit H/D disorder, hence the calculations are performed in order equivalent structure I$h$ phase which is a √3× √3×1 supercell (space group P6$_3$cm) of the actual unit cell[6]. The required force constants were computed within the Hellman-Feynman framework, on various atoms in different configurations of a supercell with (±x, ±y, ±z) atomic displacement patterns. A supercell of (2 × 2 × 2) dimension of order equivalent cell, which consist of 96 $H_2O$ molecule has been used in the computations. An energy cut-off of 900 eV was used for plane wave expansion. The Monkhorst Pack method is used for k point generation[7] with a 4×4×4 k-point mesh was used. The van-der Waals interaction has been included using the vdw-DFT method[8]. The valence electronic configurations of H and O, as used in calculations for pseudo-potential generation are $1s^1$ and $s^2p^4$ respectively. The convergence breakdown criteria for the total energy and ionic loops were set to $10^{-8}$ eV and $10^{-4}$ eV Å$^{-1}$, respectively. We have used PHONON software[9] to obtained the phonon frequencies in the entire Brillouin zone, as a subsequent step to density functional theory total energy calculations using the VASP[4, 5, 10] software. The calculation of $D_2O$ compound has been carried out using the same pseudo potential as used in $H_2O$ compound.

The thermal expansion behaviour has been computed under the quasiharmonic approximation. Each phonon mode of energy $E_{qj}$($j^{th}$ phonon mode at point q in the Brillouin zone) contributes to the thermal expansion coefficient, which is given by following relation for a hexagonal system:

$$\alpha_a(T) = \frac{1}{V_0}\sum_{q,j} C_v(q,j,T) [s_{11}\Gamma_a + s_{12}\Gamma_b + s_{13}\Gamma_c] \quad (3)$$

$$\alpha_c(T) = \frac{1}{V_0}\sum_{q,j} C_v(q,j,T) [s_{31}\Gamma_a + s_{32}\Gamma_b + s_{33}\Gamma_c] \quad (4)$$

Where $V_0$ is the unit cell volume, $\Gamma_a, \Gamma_b$ and $\Gamma_c$ are the anisotropic mode Grüneisenparameters. In a hexagonal system, Grüneisenparameters $\Gamma_a = \Gamma_b$. The mode Grüneisen parameter of phonon energy $E_{q,j}$ is given as[11],

$$\Gamma_l(E_{q,j}) = -\left(\frac{\partial \ln E_{q,j}}{\partial \ln l}\right)_{l'}; \quad l, l' = a, c \quad (5)$$

Here $s_{ij}$ are elements of elastic compliances matrix $s = C^{-1}$. $C_v(q,j,T)$ is the specific-heat contribution of the phonons of energy $E_{q,j}$.

The volume thermal expansion coefficient for a hexagonal system is given by:

$$\alpha_V = (2\alpha_a + \alpha_c) \quad (6)$$

**Phonon spectrum and partial density of states in I$h$ phase of $H_2O$ and $D_2O$**

To see the individual atomic contribution to the phonon frequency and thermal expansion behaviour, we have calculated the phonon density of states contributed from individual atoms(**Fig S4**). We learn that the translational modes of $H_2O$/$D_2O$ contribute largely at low energy regime below 40 meV. This is the reason that the thermal expansion behaviour in both the compounds at low temperature(<60K) show very similar behaviour. However due to the lighter mass of H/D than oxygen, the librational contribution of $H_2O/D_2O$ dominates at higher energy above 45 meV. The high energy spectrum in $H_2O$ at about ~225meV and ~400meV arises from the bending and stretching modes of $H_2O$. Further in $D_2O$ these bending and stretching modes are shifted to much lower energy at about 150meV and 300 meV respectively. This huge shift in energy is mainly due to the difference in mass of H and D in O-H and O-D bonds in $H_2O$ and $D_2O$ respectively.

**Anisotropic Gruneisen parameter**

In order to calculate the thermal expansion behaviour, we need to have Grüniesen parameters as derived from pressure dependence of phonon frequencies (**Fig S5**), which clearly shows that the modes below 12 meV show softening of phonon frequencies on anisotropic compression along 'a-' and 'c-'. However anisotropic compression along 'c-' axis shows that phonon modes below 2 meV have positive Grüneisen parameters ($\Gamma_c$). The behaviour of linear expansion coefficients ($\alpha_a$ and $\alpha_c$) are related to anisotropic Grüneisen parameters ($\Gamma_a$ and $\Gamma_c$) as well as elastic compliance matrix(Equations (3) and (4)). The elastic compliance matrix shows(**Table SI**) that compressibilities along 'a-' and 'c-' axis are very similar, also the anisotropic Grüneisen parameter behaivour is very similar ($\Gamma a$ and $\Gamma c$). Hence only the negative magnitude of the Grüneisen parameter will give rise to negative expansion behaviour at low temperature.

**Mean square displacement of various atom as a function of phonon energy**

The nature of dynamics of phonon modes which contributes to NTE behaviour in the compound can be understood by calculating the mean square displacement of various atom as a function of phonon energy (**Fig S7**). We can see the magnitude of mean square displacements of both the atoms (H/Dand O) are equal for the modes upto 40 meV. This suggest that phonon dynamics at low energy involves translational motion of the $H_2O$ as a rigid body and not any librations or internal modes of $H_2O/D_2O$. Hence internal dynamics of $H_2O/D_2O$ does not play any role in NTE behaviour.

**High Pressure Stability of I*h* Phase of Ice**

Ice shows large number of phase transition with pressure as well as temperature. These phase transition are expected to be related to phonon instabilities and elastic instabilities with pressure and temperature. The temperature and pressure dependence measurement of phonon modes can be used to find the soft phonon modes, which could lead to phase transition. We have calculated the pressure dependence of various Born stability criteria (Table SII). It is interesting to see that with increase in pressure the Born stability criteria related to $C_{44}$ and $C_{66}$ elastic constant are found to be violated at about 2.5 GPa. These elastic constants are related with long-wavelength transverse acoustic phonons along (100) and (110) directions in the compounds. As mentioned above,the negative thermal expansion in these compounds are attributed to transverse acoutic phonons and same phonon softening at high pressure would lead to elastic instability.

**TABLE SI**. The calculated elastic properties of the I$h$ phase of H$_2$O ice as a function of pressure.

| Compliance | 0 kbar | 10 kbar | 20 kbar | 25 kbar |
|---|---|---|---|---|
| $s_{11}$ (10$^{-4}$ kbar$^{-1}$) | 77.6 | 86.2 | 105.3 | 122.0 |
| $s_{33}$ (10$^{-4}$ kbar$^{-1}$) | 60.6 | 62.3 | 67.8 | 71.8 |
| $s_{44}$ (10$^{-4}$ kbar$^{-1}$) | 249.1 | 291.0 | 371.7 | 443.0 |
| $s_{12}$ (10$^{-4}$ kbar$^{-1}$) | -34.3 | -44.1 | -61.0 | -75.0 |
| $s_{13}$ (10$^{-4}$ kbar$^{-1}$) | -18.5 | -22.6 | -27.6 | -30.8 |
| $K_a$ (10$^{-4}$ kbar$^{-1}$) | 24.8 | 19.5 | 16.7 | 14.2 |
| $K_c$ (10$^{-4}$ kbar$^{-1}$) | 23.6 | 17.1 | 12.6 | 10.2 |
| $K$ (10$^{-4}$ kbar$^{-1}$) | 73.2 | 56.1 | 46.0 | 38.6 |
| $B$ (kbar) | 137 | 178 | 217 | 259 |

**TABLE SII**. The calculated Born stability criteria in the I$h$ phase of H$_2$O ice as a function of pressure. For an elastically stable hexagonal crystal all three Born criteria (elastic constants equations) must be positive.

| P (kbar) | Born Stability Criteria | | |
|---|---|---|---|
| | ($C_{44}$-P) (kbar) | ($C_{66}$-P) (kbar) | ($C_{33}$-P)($C_{11}$+$C_{12}$)-2($C_{13}$+P)$^2$ (kbar$^2$) |
| 0 | 40.0 | 45 | 51439 |
| 10 | 24.5 | 28.5 | 52883 |
| 20 | 7 | 10 | 42821 |
| 25 | -2.5 | 0.2 | 33335 |

Fig S1. The low energy part of experimental phonon spectra, in the I$h$ phase of $D_2O$ ice, as a function of pressure. For better visibility the spectrum from 4 to 7 meV are shown in inset.

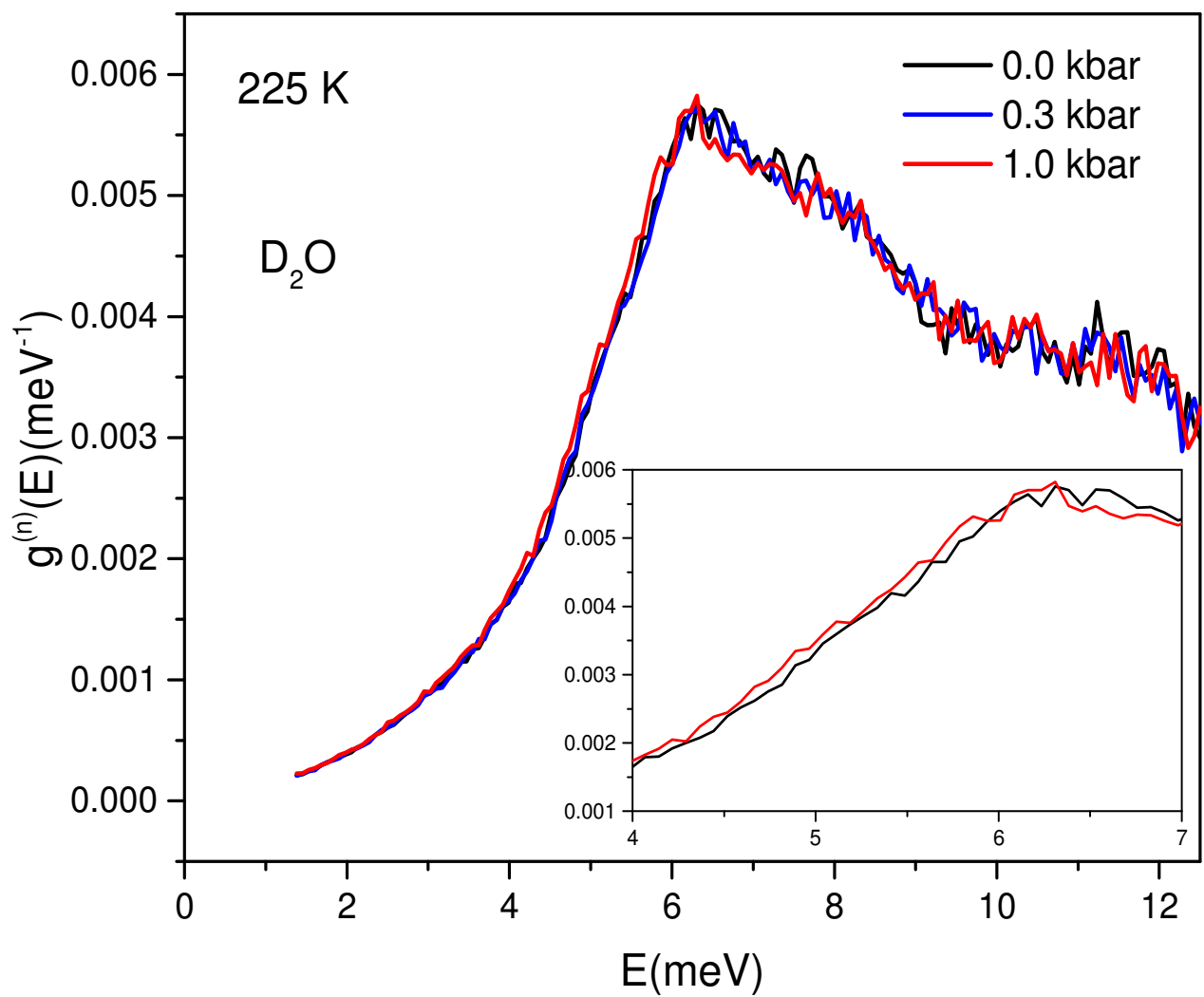

FIG S2. The calculated contribution to the linear thermal expansion coefficient from phonons of energy E averaged over entire Brillouin zone at 50K and 100K.

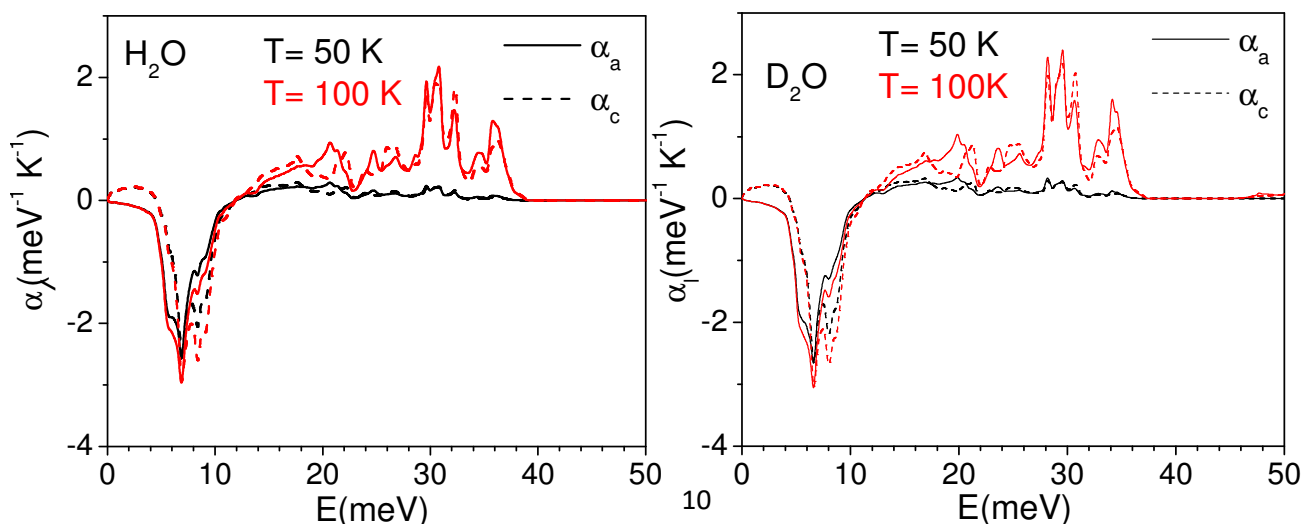

Fig S3. The calculated pressure dependence of the phonon dispersion of $H_2O$ and $D_2O$ ice in I$h$ phase. For better visibility the phonon energies are shown up to 20 meV.

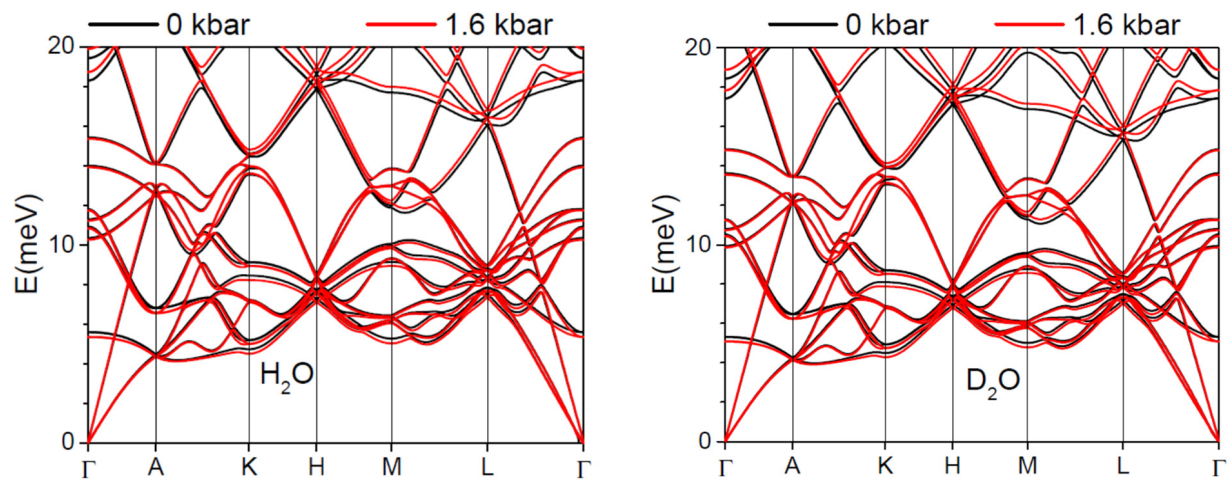

Fig S4. The calculated total and partial phonon density of states in the I$h$ phase of $H_2O$ and $D_2O$ ice.

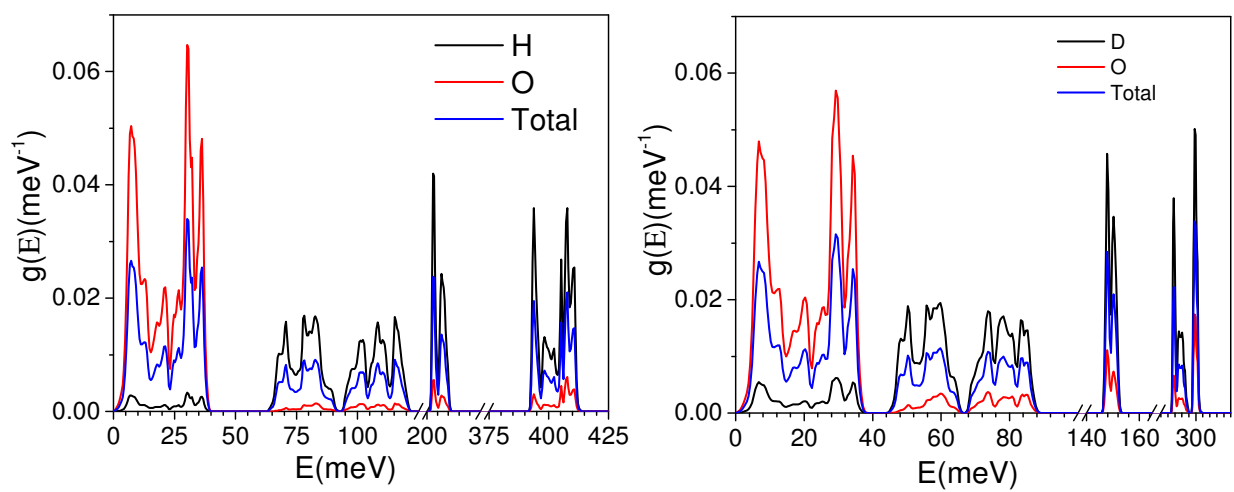

Fig S5. The calculated anisotropic Grüneisen parameters in the I$h$ phase of $H_2O$ and $D_2O$ ice.

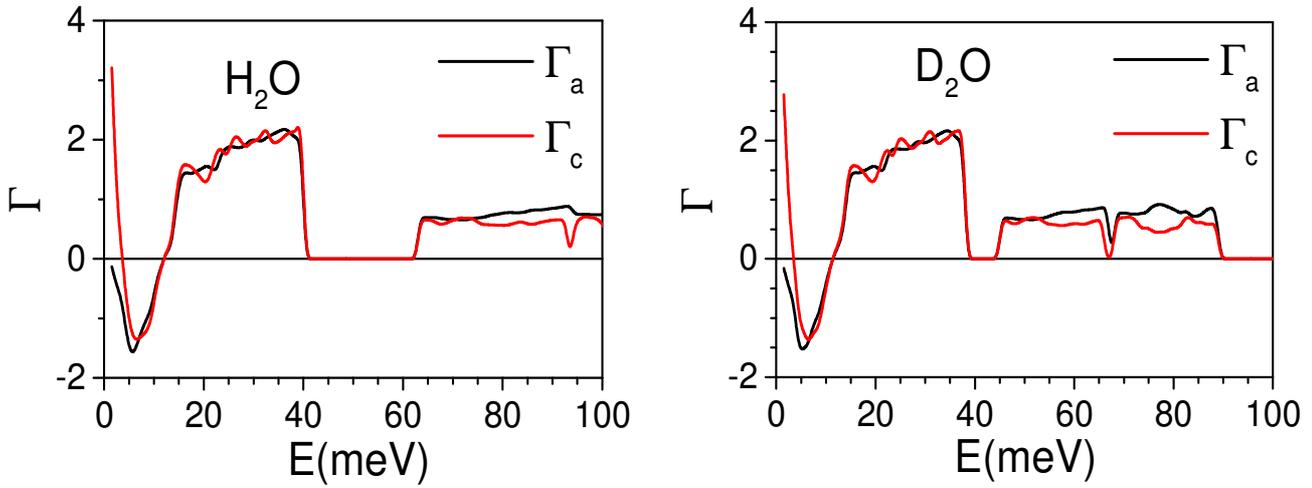

Fig S6. The calculated linear and volume thermal expansion coefficients the I$h$ phase of $H_2O$ and $D_2O$ ice.

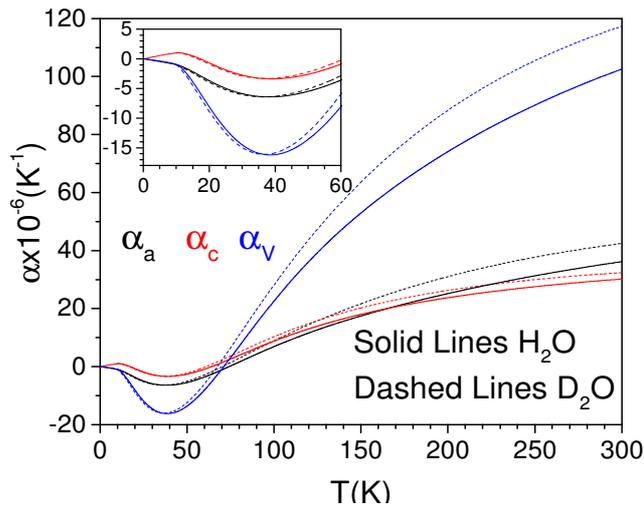

Fig S7. The calculated mean square displacement as a function of phonon energy averaged over Brillouin zone.

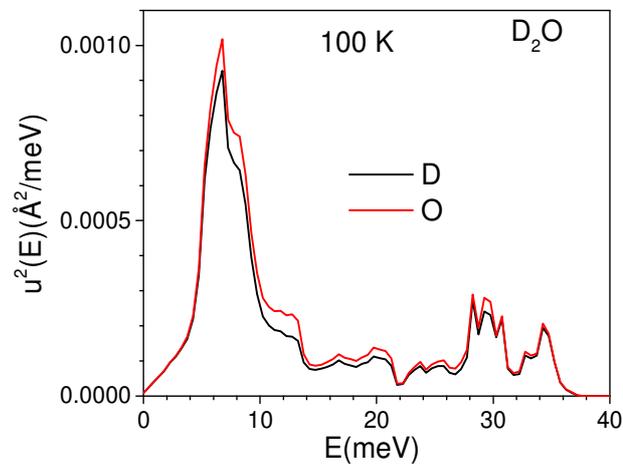